\documentclass[aps,pra,twocolumn,showpacs,superscriptaddress]{revtex4}

\usepackage{bbm}
\usepackage{graphicx}

\newcommand{\id}{I}
\def\ket#1{| #1 \rangle}

\def\kb#1#2{|#1\rangle\!\langle #2 |}

\newcommand{\ch}[1]{\mathcal{#1}}
\newcommand{\ECH}{\ch{E}}
\newcommand{\RCH}{\ch{R}}
\newcommand{\UCH}{\ch{U}}
\newcommand{\hil}{\mathcal{H}}

\newcommand{\code}{\mathcal{C}}
\newcommand{\HA}{\hil^{A}}
\newcommand{\HB}{\hil^{B}}
\newcommand{\Tr}[1]{\textrm{Tr}\left(#1\right)}

\begin{document}

\title{Optical Implementation of a Unitarily Correctable Code}

\author{K.M. Schreiter}

\affiliation{Institute for Quantum Computing, University of Waterloo, Waterloo, Ontario, N2L 3G1, Canada}
\affiliation{Department of Physics \& Astronomy, University of Waterloo, Waterloo, Ontario, N2L 3G1, Canada}

\author{A. Pasieka}

\affiliation{Department of Physics, University of Guelph, Guelph, Ontario, N1G 2W1, Canada}

\author{R. Kaltenbaek}

\affiliation{Institute for Quantum Computing, University of Waterloo, Waterloo, Ontario, N2L 3G1, Canada}
\affiliation{Department of Physics \& Astronomy, University of Waterloo, Waterloo, Ontario, N2L 3G1, Canada}

\author{K.J. Resch}

\affiliation{Institute for Quantum Computing, University of Waterloo, Waterloo, Ontario, N2L 3G1, Canada}
\affiliation{Department of Physics \& Astronomy, University of Waterloo, Waterloo, Ontario, N2L 3G1, Canada}

\author{D.W. Kribs}

\affiliation{Institute for Quantum Computing, University of Waterloo, Waterloo, Ontario, N2L 3G1, Canada}
\affiliation{Department of Mathematics \& Statistics, University of Guelph, Guelph, Ontario, N1G 2W1, Canada}

\date{\today}

\begin{abstract}

Noise poses a challenge for any real-world implementation in
quantum information science. The theory of quantum error correction deals
with this problem via methods to encode and recover quantum
information in a way that is resilient against that noise. Unitarily
correctable codes are an error correction technique wherein a single unitary recovery operation is applied without the need for an ancilla Hilbert space.
Here, we present the first optical implementation of a non-trivial
unitarily correctable code for a noisy quantum channel with no decoherence-free subspaces or
noiseless subsystems. We show that recovery of our initial states is achieved with high fidelity ($\ge0.97$), quantitatively
proving the efficacy of this unitarily correctable code.
\end{abstract}


\maketitle

\section{Introduction}

In the realm of quantum information science one of the most
pervasive obstacles that must be overcome is the effect of
interactions with the environment -- the resulting noise must be
dealt with in any quantum system of practical
interest~\cite{NC00,Got06}.  The theory of quantum error correction
has developed with this primary motivation in mind. The well-known
passive error correction (or error avoidance) notions of decoherence-free subspaces
(DFS) and noiseless subsystems
(NS)~\cite{PSE96,DG97,ZR97c,LCW98a,KLV00a,Zan01b,KBLW01} have been
extensively explored.  Experiments have demonstrated
the principle of DFSs in optical systems~\cite{KBAW00,MLRS03,BDWR04}, ion traps~\cite{KMRSIMW01}, and NMR~\cite{OLK03}. The primary appeal of passive quantum error correction schemes is the fact they \textit{are} passive --
no correction operation is required once the noise acts.  However, this
occurs at the cost of a limited number of possible correctable codes
for a given channel by relying heavily on symmetry in the
noise~\cite{CK06}.

When DFS/NS are not available or ideal, active quantum error correction (where a correction operation other
than the identity map is needed) must be employed, resulting in an expanded set of
correctable codes.  Approaches such as the stabilizer
formalism~\cite{Got96} provide formulas for finding and implementing
codes in this regime.  However, there can be a range of costs here that may negatively impact practical implementations.
Clearly then, there are benefits to be had in the identification of
correction schemes that require the application of a recovery
operation, but retain some of the appealing properties of passive codes.

Here we experimentally implement a non-trivial example of such a
correction scheme by applying random, anticorrelated noise to a two-qubit
optical system.  This noisy channel leads to the loss of encoded information and has no passive codes. However,
we show that one qubit can be encoded against the noise such that the encoded states are recovered with high fidelity by applying a single unitary recovery operation after each application of the noisy channel.  This form of \emph{unitarily
correctable code} was recently introduced~\cite{KLPL05,KS06,KPZ08}.

\section{Unitarily Correctable Codes}

One of the most general theoretical descriptions of how to deal with
environmental interactions is given by the formalism of operator
quantum error correction~\cite{KLP05,KLPL05}.  In this theory, encoding,
recovery and decoding of information takes place on subsystems of
the full system.  The key notion in operator quantum error
correction is formulated as follows:  a subsystem $\HA$ of
$\hil=\left(\HA\otimes\HB\right)\oplus\ch{K}$ is \emph{correctable}
for a quantum channel $\ECH$ if there exists a recovery operation
$\RCH$ such that for every pair of density operators $\rho^{A}\in\HA$ and
$\sigma^{B}\in\HB$ there is some other density operator
$\tau^{B}\in\HB$ such that
\begin{equation}\label{OQEC}\RCH\circ\ECH\circ\ch{P}_{AB}(\rho^{A}\otimes\sigma^{B})=\ch{P}_{AB}(\rho^{A}\otimes\tau^{B}).\end{equation}
Here $\HB$ is an ancillary subsystem, $\ch{K}$ is an ancillary subspace orthogonal to $\HA\otimes\HB$, $\ch{P}_{AB}$ is the projection superoperator of
$\hil$ onto $\HA\otimes\HB$ and by a quantum channel, we refer to a
completely-positive trace preserving map~\cite{NC00}.  Simply stated, a subsystem $\HA$ is correctable if, when we restrict our attention to it and anything it interacts with ($\HA\otimes\HB$), there is a recovery operation that returns $\HA$ to its initial state, while doing anything to $\HB$.

Decoherence-free subspaces and noiseless subsystems are two special
cases of Eq.~(\ref{OQEC}), namely when the recovery operation is the
identity map, and when $\dim{\HB}=1$ or $\dim{\HB}\geq 1$
respectively.  Active schemes such as the stabilizer formalism are examples
of Eq.~(\ref{OQEC}) in generality.

We say that a subsystem $\HA$ is \emph{unitarily correctable} for a
quantum channel $\ECH$ if there exists a unitary recovery
operation $\UCH$ such that for every pair of density operators
$\rho^{A}\in\HA$ and $\sigma^{B}\in\HB$ there is some other density
operator $\tau^{B}\in\HB$ such that
\begin{equation}\label{UCC}\UCH\circ\ECH\circ\ch{P}_{AB}(\rho^{A}\otimes\sigma^{B})=\ch{P}_{AB}(\rho^{A}\otimes\tau^{B}),\end{equation}
where $\UCH$ is a unitary map such that $\UCH:\rho\mapsto U\rho
U^{\dagger}$, with $U$ a unitary operator on $\HA\otimes\HB$.  In the general
case of Eq.~(\ref{OQEC}), every recovery operation $\RCH$ can be
implemented via a unitary map on a (in general) larger Hilbert
space~\cite{NC00}.  Thus from another perspective, unitarily
correctable codes are precisely the codes for which an extended
Hilbert space is not required in the recovery process.

It may not be immediately apparent why this is advantageous as it is
simply a special case of Eq.~(\ref{OQEC}).  The first point to
recognize is that Eq.~(\ref{UCC}) is a generalization of the DFS/NS case since
the identity map is a unitary map -- thus the number of correction
options for a given quantum channel will be at least as large.  More importantly perhaps, quantum channels with \textit{no} passive codes can gain \textit{nearly}-passive correction options.
Secondly, this potential increase in correction options comes only
at the cost of applying a single unitary map after the noise and no
measurements.  Thanks to these two points, unitarily correctable codes take advantage of two of the most appealing aspects of both active and passive correction schemes -- increased range of correction options and easily implemented correction. 

From a practical standpoint, a correction
scheme is only useful if one can find codes for which the scheme is capable of correcting a class of error operations of interest.  In the case of unitarily correctable codes this can be achieved for the class of unital operations (those for which the identity
operator is unaffected by the noise $\ECH(\id)=\id$) as shown by Theorem 2 of~\cite{KS06}.  The theorem states that the unitarily correctable codes for a quantum channel $\ECH$ are precisely the DFS/NS for $\ECH^{\dagger}\circ\ECH$,
where $\ECH^{\dagger}$ denotes the dual map for $\ECH$, defined via
the relation between expectation values: $\Tr{\ECH(\rho)X}=\Tr{\rho\ECH^{\dagger}(X)}$.
One can readily check that $\ECH$ is unital and trace preserving if and only
if $\ECH^{\dagger}$ is as well.

The problem of finding the DFS/NS for a given channel has been fully
characterized~\cite{K03,CK06}.  In the case of a unital channel, the
DFS/NS are the subsystems of the full Hilbert space that commute
with each of the Kraus operators of the channel. This fact, together with Theorem 2 of~\cite{KS06} provides a way to compute the
unitarily correctable codes for a given unital channel.
The next section includes an explicit calculation for a specific example.

\section{Experimental Model}

As a demonstrative example, we seek a simple quantum channel with
no DFS/NS but with a unitarily correctable code.  To that end,
consider the two-qubit phase-flip channel with Kraus operators
$\left\{\frac{1}{\sqrt{2}}Z_{1},\frac{1}{\sqrt{2}}Z_{2}\right\}$
where $Z_{1}=\sigma_{z}\otimes\id$ and $Z_{2}=\id\otimes
\sigma_{z}$.

First note that there are no DFS/NS for this channel:  $\ECH$ is a
unital map so we must consider the operators that commute with both
$Z_{1}$ and $Z_{2}$. In the standard basis,
$$Z_{1}=\left(\begin{array}{cccc}1&0&0&0\\0&1&0&0\\0&0&-1&0\\0&0&0&-1\end{array}\right)$$
and,
$$Z_{2}=\left(\begin{array}{cccc}1&0&0&0\\0&-1&0&0\\0&0&1&0\\0&0&0&-1\end{array}\right).$$
Considering an arbitrary matrix in the Hilbert space, $M=\left(m_{i,j}\right)$:
\begin{equation}\left[Z_{1},M\right]\\=\left(\begin{array}{cccc}0&0&2m_{1,3}&2m_{1,4}\\0&0&2m_{2,3}&2m_{2,4}\\-2m_{3,1}&-2m_{3,2}&0&0\\-2m_{4,1}&-2m_{4,2}&0&0\end{array}\right).\nonumber\end{equation}
The off-diagonal blocks must thus be equal to zero for any element of the Hilbert space to commute with $Z_{1}$ and so all elements of the Hilbert space that commute with $Z_{1}$ must be of the form (for arbitrary $a$ through $h$):
$$\left(\begin{array}{cccc}a&b&0&0\\c&d&0&0\\0&0&e&f\\0&0&g&h\end{array}\right).$$
Similarly, the elements that commute with $Z_{2}$ are:
$$\left(\begin{array}{cccc}j&0&k&0\\0&l&0&m\\n&0&o&0\\0&p&0&q\end{array}\right),$$ for arbitrary $j$ through $q$.  Therefore, the only operators that commute with both $Z_{1}$ and $Z_{2}$ are of the form $$\left(\begin{array}{cccc}r&0&0&0\\0&s&0&0\\0&0&t&0\\0&0&0&u\end{array}\right),$$ for arbitrary $r$ through $u$.  Since there are no non-zero off-diagonal elements we cannot encode any quantum information -- there are no DFS/NS for $\ECH$.

However, because $\ECH$ is a unital map we can find the unitarily
correctable codes for $\ECH$ by looking at the DFS/NS for
$\ECH^{\dagger}\circ\ECH$.  The Kraus operators for the unital
channel $\ECH^{\dagger}\circ\ECH$ are
$\left\{\frac{1}{\sqrt{2}}\id,\frac{1}{\sqrt{2}}Z_{1}Z_{2}\right\}$.
In the standard basis
$$Z_{1}Z_{2}=\left(\begin{array}{cccc}1&0&0&0\\0&-1&0&0\\0&0&-1&0\\0&0&0&1\end{array}\right),$$
thus the operators that commute with the Kraus operators of $\ECH^{\dagger}\circ\ECH$ are of the
form $$\left( \begin {array}{cccc}
a&0&0&b\\0&c&d&0\\0&e&f&0\\g&0&0&h\end {array} \right),$$ for
arbitrary $a$ through $h$.  So, there are two one-qubit unitarily
correctable codes for $\ECH$ (the two one-qubit decoherence-free
subspaces for $\ECH^{\dagger}\circ\ECH$):
$\code_{1}=\text{span}(\ket{00},\ket{11})$ and
$\code_{2}=\text{span}(\ket{01},\ket{10})$.

Finally, in order to find a suitable correction operation, consider
the effect of $\ECH$ on an arbitrary two-qubit density matrix
$\rho=\left(\rho_{i,j}\right)$:
\begin{equation}\label{channel_effect}\ECH(\rho)=\left( \begin
{array}{cccc}
\rho_{1,1}&0&0&-\rho_{1,4}\\0&\rho_{2,2}&-\rho_{2,3}&0\\0&-\rho_{3,2}&\rho_{3,3}&0\\-\rho_{4,1}&0&0&\rho_{4,4}\end
{array} \right).\end{equation}  There are many candidate unitary
correction operations -- for $\code_{1}$ the controlled-phase gate
is one -- it is also easy to see that both $Z_{1}$ and $Z_{2}$ will
correct either of $\code_{1}$ or $\code_{2}$ when taken as the
unitary matrix of Eq.~(\ref{UCC}).  In any experimental setting, some gates are easier to implement than others (e.g. in the optical setting, 2-qubit gates are difficult because of the lack of photon-photon interactions).  The flexibility in the choice of the unitary correction operation is another appealing aspect of unitarily correctable codes.

We have found what we set out to look for, a quantum channel with
no DFS/NS, but two unitarily correctable codes.  Although one qubit
cannot be encoded with no recovery operation, one can be encoded
with a single unitary recovery.  Therefore this model allows for a
clear experimental demonstration of one of the key advantages of the unitarily
correctable code approach -- we gain a \textit{nearly-passive correction
scheme} for a quantum channel with \textit{no passive codes}.

An additional advantage of this model is that it can be realized
in an optical setting.  Defining $\ket{H}\equiv\ket{0}$ and
$\ket{V}\equiv\ket{1}$, a phase-flip can be achieved by placing a
half-wave plate (HWP) with its optic axis oriented
along $\ket{H}$ in the path of a photon.  Thus $Z_{1}$ ($Z_{2}$) is implemented by a HWP in
the first (second) photon path with nothing in the second (first).
Since both $Z_{1}$ and $Z_{2}$ will correct either codespace,
we can experimentally implement the correction by placing a HWP in either photon path.

Thus, the particular optical implementation of the unitarily
correctable code we consider is as follows: considering
$$\UCH\circ\ECH\circ\ch{P}_{AB}(\rho^{A}\otimes\sigma^{B})=\ch{P}_{AB}(\rho^{A}\otimes\tau^{B}),$$
the full Hilbert space $\hil$ is the set of polarization states of
two entangled photons.  $\hil$ can be decomposed as
$\hil=\left(\HA\otimes\HB\right)\oplus\ch{K}$ where $\dim{\HB}=1$.
The single encoded qubit will be supported on
$P_{\code_{1}}=\kb{00}{00}+\kb{11}{11}$, so $\dim{\HA}=2$, and the
ancilla space $\ch{K}$ is supported on
$P_{\code_{1}^{\perp}}=\kb{01}{01}+\kb{10}{10}$ and also has dimension
$2$.  Thus the projection superoperator $\ch{P}_{AB}$ focuses our
attention onto the codespace $\code_{1}$: $\ch{P}_{AB}:\rho\mapsto P_{\code_{1}}\rho
P_{\code_{1}}$.  The noise is provided by randomly fired anti-correlated phase-flips in both
photon paths and has the form $\ECH:\rho\mapsto\frac{1}{2}Z_{1}\rho
Z_{1}+\frac{1}{2}Z_{2}\rho Z_{2}$.  Finally, the correction
operation is provided by a single HWP placed in the second photon path,
set to apply $Z_{2}$, so $\UCH:\rho\mapsto Z_{2}\rho Z_{2}$.

\section{Experimental Setup}

\begin{figure}[h!]
\begin{center}
\includegraphics[width=1 \columnwidth]{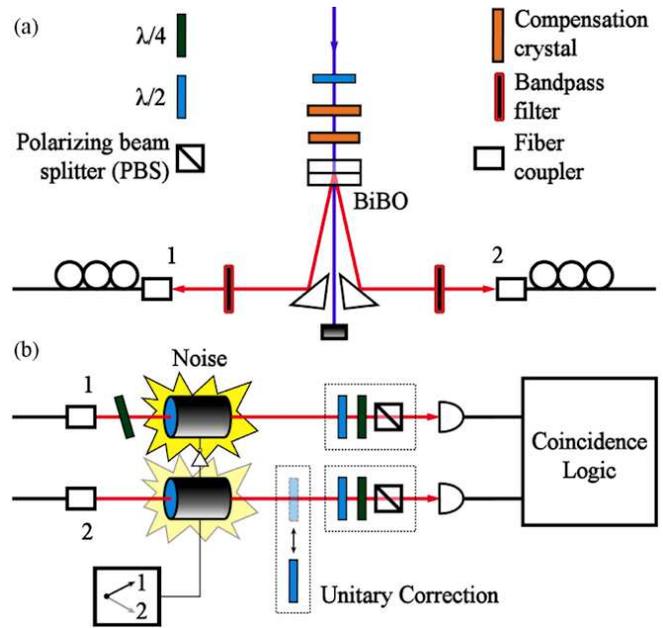}
\caption{(Color online) Experimental setup.  (a) Source of polarization-entangled
photon pairs. An ultraviolet diode laser is used to pump a pair of
orthogonally-oriented BiBO crystals to produce entangled photon
pairs by spontaneous parametric down-conversion.  The precise state produced is
determined by the polarization of the laser light and can be set
using a half-wave plate (HWP).  A pair of compensation crystals are
used to compensate group-velocity mismatch in the down-conversion crystals. The
light is collected into single-mode fibers.  More details can be
found in the text. (b)  Experimental implementation of the noise
model, correction, and tomography. A quarter-wave plate is used to
adjust the phase of the entangled pairs; together with the HWP in
the UV laser and the fiber-based polarization controllers, these operations are sufficient to prepare an
arbitrary pure state in $\code_{1}$.  When the noise is on, the photon
pairs are subject to computer controlled anti-correlated phase
errors through liquid-crystal variable phase retarders (LCVR); a
decision as to which LCVR will fire is made by a pseudo-random
number generator every 1~s.  The state of the light is measured using
quantum state tomography with a tomographically overcomplete set of
measurements and the maximum likelihood procedure \cite{James01}.
\label{experiment}}
\end{center}
\end{figure}

\begin{figure}[t!]
\begin{center}
\includegraphics[width=0.9 \columnwidth]{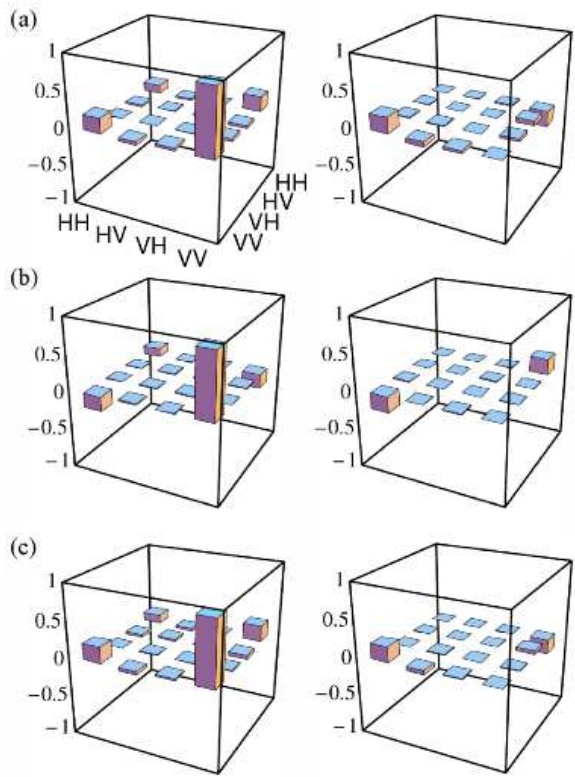}
\caption{(Color online) Example set of experimentally measured two-photon density
matrices (left: real parts, right: imaginary parts). (a) The initial state was chosen to have a non-trivial
value of both angles in Eq.~(\ref{generalcodestate}); the nearest pure state in $\code_{1}$ has angles %$\theta=71.0^\circ$ 
$\theta=35.5^\circ$ and $\phi=46.5^\circ$.  (b) The
noise affected state has its fidelity to the initial state
reduced to 0.62; one can see that the primary effect of the noise is
to flip the sign of the real and imaginary parts of the coherences
rather than decohere the state.  (c) The final state after both noise
and correction.  The fidelity with the initial state has been
restored to $0.97$ thus demonstrating the efficacy of the
unitarily correctable code. \label{sample2dms}}
\end{center}
\end{figure}

Our experimental source is shown in Fig.~\ref{experiment}(a).  We use
a 185~mW free-running UV diode laser (Newport model: LQC405-180E,
centre wavelength $404.4$~nm, bandwidth 0.8~nm FWHM). To improve mode quality, the beam is passed through a
spatial filter to obtain a 55~mW near-TEM$_{00}$ beam.  A motorized
half-wave plate then rotates the pump polarization from horizontal
to an arbitrary linear polarization. The diode laser pumps a pair of
0.5~mm thick orthogonally-oriented BiBO nonlinear crystals cut for
type-I non-collinear degenerate down-conversion with an 3$^\circ$
half-opening angle outside the crystal \cite{Kwiat99}.  A 1~mm
$\alpha$-BBO and a 1~mm quartz both cut for maximum birefringence
are used to compensate group-velocity mismatch in the down-conversion
crystals. Entangled photon pairs emitted by the source pass through
bandpass filters (centre wavelength 810~nm, bandwidth 5~nm FWHM) and
are coupled into single-mode optical fibers.

The light is coupled back into free space before application of the
noise, as shown in Fig.~\ref{experiment}(b). The noise $\ECH$ is
implemented by a pair of computer controlled liquid-crystal variable
phase retarders (Meadowlark LVC-100, LCVR), one in each photon path.
The LCVRs are both set to implement
      phase-flips. To simulate a noisy channel \textit{either} one or
      the other LCVR is randomly fired with a switching rate of $1\;$Hz
      providing anti-correlated noise. The random switching was
      implemented using a software pseudo-random number generator
      (LabView, National Instruments).  Recovery is achieved
with a half-wave plate oriented along $\ket{H}$ in the path of
photon 2.

We characterize all of our states using quantum state tomography,
Fig~\ref{experiment}(b).  There is a polarization analyzer in each
arm comprised of a half-wave plate, a quarter-wave plate, and a
polarizing beam-splitter.  We use the tomographically-overcomplete
polarization measurements $\{H,V,D,A,R,L\}$ for each photon ($R$ and
$L$ represent right- and left-circular polarizations), i.e., 36
measurement settings on the pair of photons. Density matrices are
reconstructed from the experimentally measured counts using the
maximum-likelihood method \cite{James01}. We performed quantum state
tomography at each of three stages in the experiment: in the
absence of noise or correction, in the presence of noise, and in the presence of both noise and correction.

Typical rates for the source are approximately 12000 coincidence
counts/s and 60000 singles counts/s when the fibers are directly
connected to detectors.  Using the half-wave plate placed before the
down-conversion crystals, the fiber-based polarization controllers,
and the tilted quarter-wave plate in the path of photon 1, the
source can produce output states in $\code_{1}$ of the form
\begin{equation}
\ket{\psi}=\cos2\theta \ket{HH} + \sin2\theta e^{i\phi} \ket{VV}.
\label{generalcodestate}
\end{equation}
With the half-wave plate oriented at
$\theta=22.5^\circ$ and the phase angle at $\phi=0$ the system ideally produces the Bell state
$\ket{\phi^+}=\frac{1}{\sqrt{2}}(\ket{HH}+\ket{VV})$. In this
case, we measure visibilities of 99.2\% in the horizontal/vertical
(H/V) basis and 95.3\% in the diagonal/antidiagonal (D/A) basis.
Reconstructing the state using quantum state tomography we find, in
this case, that the fidelity~\cite{J04,FN01} with $\ket{\phi^{+}}$ is $0.97$, the tangle~\cite{CKW00}
is $\tau=0.9$, and the linear entropy~\cite{FN02} is
$S_{L}=0.065$.  Thus our source is capable of producing highly entangled and
nearly pure states.

\section{Results \& Discussion}

\begin{figure}[t!]
\begin{center}
\includegraphics[width=1 \columnwidth]{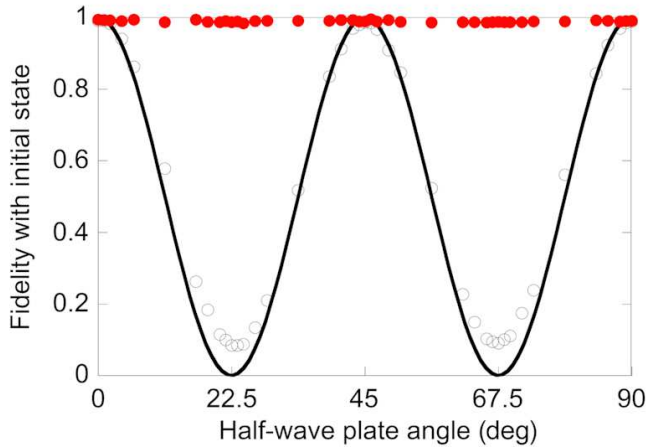}
\caption{(Color online) Measured fidelity between initial and final states. The
mixed state fidelity \cite{J04} was calculated on
experimentally reconstructed density matrices between the initial
states and noise-affected states (open circles, solid line theory)
and the initial states and states affected by noise and unitary
correction (closed, red circles).  The data show that the effect of
the noise depends strongly on the initial state.  Initial states
closest to the product states $|HH\rangle$ and $|VV\rangle$ are
least affected while those close to the maximally entangled state
$|\phi^+\rangle$ are driven to nearly orthogonal states.  The
correction restores the photons to states with fidelity of higher than $0.98$ with
the initial states in all cases. 
\label{summary2qb}}
\end{center}
\end{figure}

An example set of reconstructed density matrices, chosen to show the
effect of the noise and correction on a state with unequal populations, and both real and
imaginary coherences, is shown in Fig.~\ref{sample2dms}. Data was
accumulated for 5~s per measurement setting.  The initial state is
shown in Fig.~\ref{sample2dms}(a). This state has fidelity $0.98$ with
a state in $\code_{1}$ [Eq.~(\ref{generalcodestate})] with 
$\theta=35.5^\circ$ and $\phi=46.5^\circ$.  Subjecting the photon
pairs to the noise changes the state to the one shown in
Fig.~\ref{sample2dms}(b). The fidelity of this state to the initial
state in Fig.~\ref{sample2dms}(a) has been reduced to $0.62$. Qualitatively, one can see that the effect of the
noise $\ECH$ does not decohere these states, but rather flips the
sign of both the real and imaginary coherences. The correction,
$\UCH$, is implemented by placing a half-wave plate in the path of
photon 2. When the noisy state was corrected, we obtained the state
shown in Fig.~\ref{sample2dms}(c) where the fidelity with the
initial state has been restored to $0.97$. The recovery of high
fidelity states demonstrates the effectiveness of this unitarily
correctable code.

We characterized the effect of the noise and noise plus correction
on a range of input states.  We adjusted the source to states in $\code_{1}$ with real coefficients ($\phi=0$) and tuned those
coefficients by adjusting the angle of the HWP %$\vartheta$,
in the pump laser, $\theta$.  The fidelity measurements of the noise affected states are shown in
Fig.~\ref{summary2qb} where the initial states are shown as open
circles and the theoretical prediction, $F=\text{cos}^{2}{4\theta}$, is a
solid black line. These data clearly show that the noise affects the
states, but by different amounts.  As expected, those states closer to the
product states $\ket{HH}$ and $\ket{VV}$, or HWP settings
$0^\circ$ (or $90^\circ$) and $45^\circ$, respectively, are less
affected by the noise; in the case of $\ket{HH}$ and
$\ket{VV}$, the states are invariant under the noise. Those states closer to the maximally
entangled states $\ket{\phi^+}$ and
$\ket{\phi^-}=\frac{1}{\sqrt{2}}(\ket{HH}-\ket{VV})$, or HWP
settings $22.5^\circ$ and $67.5^\circ$, respectively, are most
affected by the noise as it drives them to nearly orthogonal states.
The states are restored to greater than $0.98$ fidelity in all measurements thereby demonstrating the effectiveness of this unitarily correctable code across the entire codespace $\code_{1}$.

\section{Conclusion}

We have described a simple noise model which does not have a
decoherence-free subspace or noiseless subsystem but does contain two unitarily correctable
codes.  We have implemented this noise model by constructing a
quantum channel in a physical system, namely the polarization of a
pair of optical photons, in which two liquid-crystal variable phase
retarders fire in an anticorrelated, but random manner.  We have
sent photon pairs through this noisy channel in states ranging from
separable to nearly maximally entangled and compared the impact of
the noise and the noise plus unitary correction on the quality of
the states.  Our data shows that the noise can dramatically impact
the fidelity of the output state with the initial, especially in the
case of highly entangled states in $\code_{1}$.  However, the unitary correction
restores the state quality with fidelities greater than $0.97$ for all states in $\code_{1}$.

Our experiment shows how to translate the theory of unitarily correctable codes, a new method combining aspects of passive
and active quantum error correction, into experimental realization.
As these codes expand the class of nearly-passive correctable codes, it is an interesting open question as to how these approaches could be utilized in more natural physical systems, as opposed to controlled
application of noise.

\section{Acknowledgments}

This work was funded by NSERC, Ontario MRI ERA, CFI, IQC, and OCE.
A.P. was partially supported by an Ontario Graduate Scholarship.
R.K. acknowledges financial support from IQC.  We thank Zhenwen Wang for technical assistance with electronics.

%\bibliography{SPKRK09}

\begin{thebibliography}{10}
\bibitem{FN01}
The mixed state fidelity is $F=\left[\rm{Tr}\sqrt{\sqrt{\rho_1}\rho_2\sqrt{\rho_1} }\right]^2$.
\bibitem{FN02}
Linear entropy is a measure of the purity of a quantum state ranging from 0 for pure states to 1 for maximally mixed states. For two qubits, it is defined $S_L=\frac{4}{3}(1-Tr\rho^2)$.
\end{thebibliography}
%\bibliographystyle{h-physrev}

\end{document}